\documentclass[useAMS,usenatbib]{mn2e}
\usepackage{txfonts,graphicx,amssymb,subfigure}
\usepackage[normalem]{ulem}
\usepackage{natbib}

\def\cm{\,\mathrm{cm}}
\def\mum{\,\mu \mathrm{m}}
\def\cmsqs{\,\mathrm{cm}^2/\mathrm{s}}

\def\kms{\,\mathrm{km\,s}^{-1}}

\def\muJy{\,\mu \mathrm{Jy}}

\def\muG{\,\mu \mathrm{G}}
\def\yr{\,\mathrm{yr}}

\def\GHz{\,\mathrm{GHz}}
\def\pc{\,\mathrm{pc}}
\def\kpc{\,\mathrm{kpc}}

\def\Btur{B_\mathrm{tur}}
\def\Bord{B_\mathrm{ord}}
\def\Btot{B_\mathrm{tot}}
\newcommand{\pheins}{\phantom{1}}

\newcommand{\phelf}{\phantom{11}}

\begin{document}

\title[Propagation of cosmic-ray electrons in M\,31 and M\,33]{How
cosmic-ray electron propagation affects radio -- far-infrared
correlations in M\,31 and M\,33}

\author[E. M. Berkhuijsen et al.]{E. M. Berkhuijsen$^{1}$
\thanks{E-mail: eberkhuijsen@mpifr-bonn.mpg.de}, R. Beck$^{1}$ and F. S.
Tabatabaei$^{2}$\\
$^{1}$MPI f\"ur Radioastronomie, Auf dem H\"ugel
69, 53121 Bonn, Germany\\ $^{2}$ MPI f\"ur Astronomie, K\"onigstuhl
17, 69117 Heidelberg, Germany}

\date{Accepted 2013 July 26. Received 2013 July 13; in original form 2013 January 18}

\maketitle

\begin{abstract}
We investigate the effect of propagation of cosmic- ray electrons
(CRE) on the nonthermal (synchrotron) -- far-infrared correlations
in M\,31 and M\,33. The thermal (TH) and nonthermal (NTH) emission
components of the radio continuum emission at 1.4~GHz and one higher
frequency are compared with dust emission from M\,31 and M\,33
using Spitzer data. In both galaxies the TH emission is linearly
correlated with the emission from warm dust $(24\mum,\ 70\mum)$, but
the power laws of the NTH -- FIR correlations have exponents $b<1$
that increase with increasing frequency. Furthermore, the values of
$b$ for M\,33 are significantly smaller ($b\simeq0.4$) than those for
M\,31 ($b\simeq0.6$). We interpret the differences in $b$ as differences
in the diffusion length of the CRE. We estimate the diffusion length
in two ways: (1) by smoothing the NTH emission at the higher frequency
until the correlation with NTH emission at 1.4~GHz has $b=1$, and (2) by
smoothing the TH emission until the correlation with the
NTH emission at the same frequency has $b=1$, assuming that the TH emission
represents the source distribution of the CRE. Our smoothing experiments
show that M\,31 only has a thin NTH disk with a scale height of
$h=0.3-0.4 \kpc$ at 1.4~GHz, whereas M\,33 has a similar thin disk as
well as a thick disk with scale height $h_\mathrm{thick}\simeq2\kpc$.
In the thin disks, the (deprojected) diffusion length at 1.4~GHz is
$\simeq 1.5 \kpc$, yielding a diffusion coefficient of $\simeq 2 \,
10^{28}\cmsqs$. The structure, strength and regularity of the magnetic
field in a galaxy as well as the existence of a thick disk determine
the diffusion of the CRE, and hence, the power-law exponent of the
NTH -- FIR correlations.
\end{abstract}



\section{Introduction}

The radio continuum and far-infrared (FIR) luminosities of
star-forming galaxies are tightly correlated over five orders of
magnitude (e.g. \cite{deJong85,Helou85,Condon92,Yun01}). The
correlation is nearly linear and probably arises because both
emissions depend on the recent star-formation rate in a galaxy.
Massive stars heat the dust and ionize the gas causing thermal dust
emission and thermal (free-free) radio continuum emission,
respectively. These stars are progenitors of supernova remnants,
which are the sources of cosmic ray electrons (CRE) that radiate
nonthermal (synchrotron) emission when spiraling around magnetic
field lines. Attempts to explain the correlation were made by
\cite{Voelk89}; \cite{Helou93}; \cite{Niklas97} and
\cite{Hoernes98}. Recently, \cite{Bell03} and \cite{Lacki10} showed
that several factors conspire in making the global correlation
linear over five decades.

The radio -- FIR correlation \footnote{Throughout this paper, the
radio and FIR intensities are on the Y- and X-axis, respectively.}
also holds within star-forming galaxies down to scales of below
hundred parsec (e.g. \cite{Beck88, Bicay89, Fitt92, Xu92, Lu96,
Hoernes98, Hippelein03, Hughes06, Paladino06, Dumas11, Leverenz12,
Tabatabaei13a}). However, many authors (starting with \cite {Xu92})
found that the radio -- FIR correlation within galaxies only
becomes linear on star-forming regions but is non-linear, usually
with power-law exponent $<1$, in other parts of a galaxy.
Separation of thermal/nonthermal radio emission and of warm/cool
dust emission revealed that the thermal radio -- warm dust
correlation is linear, whereas the nonthermal radio-cool dust
correlation has  an exponent $\not= 1$ \citep{Xu92, Hoernes98,
Hippelein03, Basu12}. This explains why the radio -- FIR
correlation of the total emission steepens on star-forming regions
where the thermal fraction of the radio emission is high
\citep{Hughes06, Paladino06, Dumas11}. The smaller exponent of the
nonthermal -- FIR correlation is probably due to diffusion of the
CRE away from their places of origin. By smoothing images of warm
dust emission until they resembled those of radio emission at
1.4~GHz propagation lengths of up to several kpc were found
(\cite{Bicay90, Marsh98, Murphy06a, Murphy06b, Murphy08, Murphy12}).

On each of our nearest spiral galaxies, M\,31 and M\,33, one radio
-- FIR study is available. The radio -- FIR correlation within M\,31
was extensively studied by \cite{Hoernes98}, who were the first to
separate thermal/nonthermal emission at 1.4~GHz and warm/cool dust
emission using HIRES data. By means of classical, pixel-to-pixel
correlations and rigorous statistics they found significant
relationships between thermal radio and warm dust emission and
between nonthermal radio and cool dust emission, with power-law
exponents of $1.2\pm0.1$ and $0.8\pm0.1$, respectively.

\cite{Hippelein03} studied the dust distribution in M\,33 observed
by ISOPHOT. They also plotted the total 4.8~GHz luminosity of
star-forming regions against that of the total FIR and obtained a
power law with exponent 0.9. The exponent is $<1$ because 30--40\%
of the 4.8~GHz emission from these regions is nonthermal
\citep{Tabatabaei07b}. Below we compare radio -- FIR correlations
in M\,31 and M\,33 using new thermal and nonthermal radio maps and
recent FIR data.

\cite{Tabatabaei07b} obtained the distribution of the thermal
emission at 21.0~cm (1425~MHz) and 3.6~cm (8350~MHz) from M\,33 from
the extinction-corrected H${\alpha}$ map of \cite{Hoopes00}. The
extinction correction was based on the optical depth map derived
from Spitzer MIPS data at $70\mum$ and $160\mum$ presented by
 \cite{Tabatabaei07a}. By subtracting the corrected thermal
emission from the total radio emission observed by \cite{Tabatabaei07a},
they obtained the distributions of the nonthermal emission across
M\,33 at 21~cm and 3.6~cm. This new method of separating
thermal/nonthermal radio emission from a galaxy is preferable above
the standard method, which requires the unrealistic assumption of
a constant nonthermal spectral index between the radio emissions at
two wavelengths.

Recently, \cite{Tabatabaei10} derived an optical depth map of M\,31
from the $70\mum$ and $160\mum$ Spitzer MIPS maps of \cite{Gordon06}
and used this to correct the H${\alpha}$ map of \cite{Devereux94}
for extinction. Taking the gradient in electron temperature into
account, we converted this map to distributions of thermal emission
at 20.5~cm (1465~MHz) and 6.3~cm (4850~MHz). Subtraction of the
thermal emission from the total emission at the corresponding
wavelength (Beck et al. 1998; Berkhuijsen et al. 2003) then yielded
the distributions of the nonthermal emission at 20.5~cm and 6.3~cm.
The details of this procedure are described in \cite{Tabatabaei13b}
where also the thermal and non-thermal maps at 1.4\,GHz of both galaxies
are shown.

In this paper we correlate the new thermal and nonthermal
distributions of M\,31 and M\,33 with the distributions of the dust
emission. We discuss the differences in the correlation results for
the two galaxies focussing on the diffusion of CRE. The correlations
and statistics are presented in Sect.~2. We derive diffusion lengths
of CRE in the plane of the sky in Sect.~3 and discuss these results
in Sect.~4. In Sect.~5 we describe the propagation of CRE within and
perpendicular to the disk of each galaxy, and we discuss the role of
star formation and magnetic fields in Sect.~6. We compare with
earlier work on CRE propagation in Sect.~7 and summarize our
conclusions in Sect.~8.

Throughout the paper, we assume a distance
$D=780\kpc$ \citep{Stanek98} and inclination $i=75\degr$ for M\,31,
and $D=840\kpc$ \citep{Freedman91} and $i=56\degr$ for M\,33.

\section{Classical correlations between radio and dust emissions}

For a fair comparison of the radio -- FIR correlations in M\,31 and
M\,33, we selected regions that contain a range of star-formation
rates and are fully covered by all data sets. This is the case for
the bright ``ring'' of emission in M\,31 in the radial range
$R=6.8\kpc - 12.5\kpc\ (30\arcmin - 55\arcmin)$ and in M\,33 at
radii $R<5\kpc\ (<20\farcm 5)$. For M\,31 we made correlations at
the angular resolution of 45\arcsec\ (corresponding to 170\,pc x
660\,pc along the major and minor axis in the plane of the galaxy)
of the 20.5~cm data and at the resolution of 3\arcmin\ (680\,pc x
2630\,pc) of the 6.3~cm data. The dust maps were smoothed to the
resolutions of the radio maps by using the custom kernels
presented by \cite{Gordon07}. All data for M\,33 were smoothed to
the resolution of 90\arcsec\ (370\,pc x 655\,pc) of the 3.6~cm data.
We note that the angular resolution does not affect the correlation
results, provided it is the same for the data sets correlated.

We made pixel-to-pixel correlations between the distributions of the
thermal emission (TH) and those of the dust emission at $24\mum ,\
70\mum$ and $160\mum$ as well as between the distributions of the
non-thermal emission (NTH) and those of the dust emission. Only
pixels with intensities above $2\times$ rms noise were used. We
obtained sets of independent data points, i.e. a beam area overlap
of $<5$\%, by choosing pixels spaced by more than $1.67\times$ the
beamwidth. Since the correlated variables are not directly depending
on each other, we fitted a power law by determining the OLS bisector
in the log-log plane in each case \citep{Isobe90}.

We also calculated the correlation coefficient, $r_\mathrm{c}$, to
show how well two components are correlated, and the student-t test
\citep{Wall79} to indicate the statistical significance of the fit.
For a number of independent points of $N >60$, the fit is
significant at the $3\sigma$ level if $t>3.1$. Errors in intercept
$a$ and slope $b$ of the bisector are standard deviations
($1\sigma$). The bisector fits of the correlations for M\,31 and
M\,33 are given in Tables~1 and 2, respectively.

\begin{table*}
\caption{M\,31: Classical correlations between FIR and radio emission
for $6.8 < R < 12.5\kpc$. OLS bisector fits with N the number of
independent data points, $r_\mathrm{c}$ the correlation coefficient
and t student-t test.} \label{tab1}
\begin{tabular}{l l c c c c c}
\hline\hline
X         &Y          &\multicolumn{2}{c}{log(Y)=a+b*log(X)} &N  &Corr. &t \\
$[$MJy/sr$]$ &$[\mu$Jy/kpc$^2]$&a          &b                &   &Coeff. & \\
          &                 &              &                 &   &$r_\mathrm{c}$ & \\
\hline
I(24)     &TH(20.5)  &4.18$\pm$0.01  &1.08$\pm$0.02   &1036  &0.77$\pm$0.02 &39 \\
I(70)     &TH(20.5)  &3.20$\pm$0.01  &0.99$\pm$0.02   &1023  &0.78$\pm$0.02 &40 \\
I(160)    &TH(20.5)  &1.91$\pm$0.04  &1.42$\pm$0.03   &1047  &0.77$\pm$0.02 &38 \\
\cline{1-7}
I(24)     &NTH(20.5) &4.62$\pm$0.01  &0.61$\pm$0.02   &1093  &0.67$\pm$0.02 &29 \\
I(70)     &NTH(20.5) &4.07$\pm$0.01  &0.56$\pm$0.01   &1079  &0.67$\pm$0.02 &30 \\
I(160)    &NTH(20.5) &3.34$\pm$0.03  &0.78$\pm$0.02   &1105  &0.67$\pm$0.02 &30 \\
\cline{1-7}
I(24)     &NTH(6.3)  &4.46$\pm$0.03  &0.97$\pm$0.09   &\phelf 65  &0.54$\pm$0.11 &\pheins 5 \\
I(70)     &NTH(6.3)  &3.56$\pm$0.06  &0.87$\pm$0.08   &\phelf 65  &0.56$\pm$0.11 &\pheins 5 \\
I(160)    &NTH(6.3)  &2.48$\pm$0.16  &1.21$\pm$0.11   &\phelf 65  &0.58$\pm$0.10 &\pheins 6 \\
\cline{1-7}
TH(20.5)  &NTH(20.5) &2.11$\pm$0.04  &0.61$\pm$0.02   &1044       &0.51$\pm$0.03 &19 \\
TH(6.3)   &NTH(6.3)  &0.93$\pm$0.27  &0.84$\pm$0.08   &\phelf 64  &0.41$\pm$0.12 &\pheins 4 \\
\hline
\multicolumn{7}{l}{Note: Correlations between TH(6.3) and dust emission agree within errors with} \\
\multicolumn{7}{l}{those between TH(20.5) and dust emission.} \\
\end{tabular}
\end{table*}

\begin{table*}
\caption{M\,33: Classical correlations between FIR and radio emission
for $R < 5\kpc$. OLS bisector fits with N the number of independent
data points, $r_\mathrm{c}$ the correlation coefficient and t
student-t test.} \label{tab2}
\begin{tabular}{l l c c c c c}
\hline\hline
X         &Y          &\multicolumn{2}{c}{log(Y)=a+b*log(X)} &N  &Corr. &t \\
$[$MJy/sr$]$ &$[\mu$Jy/kpc$^2]$&a          &b                &   &Coeff. & \\
          &                 &              &                 &   &$r_\mathrm{c}$ & \\
\hline
I(24)     &TH(21)    &4.41$\pm$0.02  &1.06$\pm$0.04   &158   &0.89$\pm$0.04   &25 \\
I(70)     &TH(21)    &3.03$\pm$0.03  &1.20$\pm$0.04   &160   &0.91$\pm$0.03   &27 \\
I(160)    &TH(21)    &2.21$\pm$0.07  &1.40$\pm$0.05   &160   &0.87$\pm$0.04   &23 \\
\cline{1-7}
I(24)     &NTH(21)   &4.62$\pm$0.01  &0.35$\pm$0.02   &159   &0.75$\pm$0.05   &14 \\
I(70)     &NTH(21)   &4.17$\pm$0.01  &0.39$\pm$0.02   &161   &0.80$\pm$0.05   &17 \\
I(160)    &NTH(21)   &3.90$\pm$0.03  &0.46$\pm$0.02   &161   &0.80$\pm$0.05   &17 \\
\cline{1-7}
I(24)     &NTH(3.6)  &4.11$\pm$0.02  &0.59$\pm$0.04   &134   &0.56$\pm$0.07   &\pheins 8 \\
I(70)     &NTH(3.6)  &3.34$\pm$0.04  &0.67$\pm$0.05   &135   &0.57$\pm$0.07   &\pheins 8 \\
I(160)    &NTH(3.6)  &2.87$\pm$0.08  &0.79$\pm$0.06   &135   &0.51$\pm$0.08   &\pheins 7 \\
\cline{1-7}
TH(21)    &NTH(21)   &2.84$\pm$0.05  &0.34$\pm$0.02   &160   &0.69$\pm$0.06   &12 \\
TH(3.6)   &NTH(3.6)  &1.33$\pm$0.15  &0.61$\pm$0.05   &134   &0.53$\pm$0.07   &\pheins 7 \\
\hline
\multicolumn{7}{l}{Note: Correlations between TH(3.6) and dust emission agree within errors with} \\
\multicolumn{7}{l}{those between TH(21) and dust emission.} \\
\end{tabular}
\end{table*}

\begin{figure*}
\centering
\vspace{0.2cm}
\begin{tabular}{cccc}
\includegraphics[width=0.40\textwidth,angle=270]{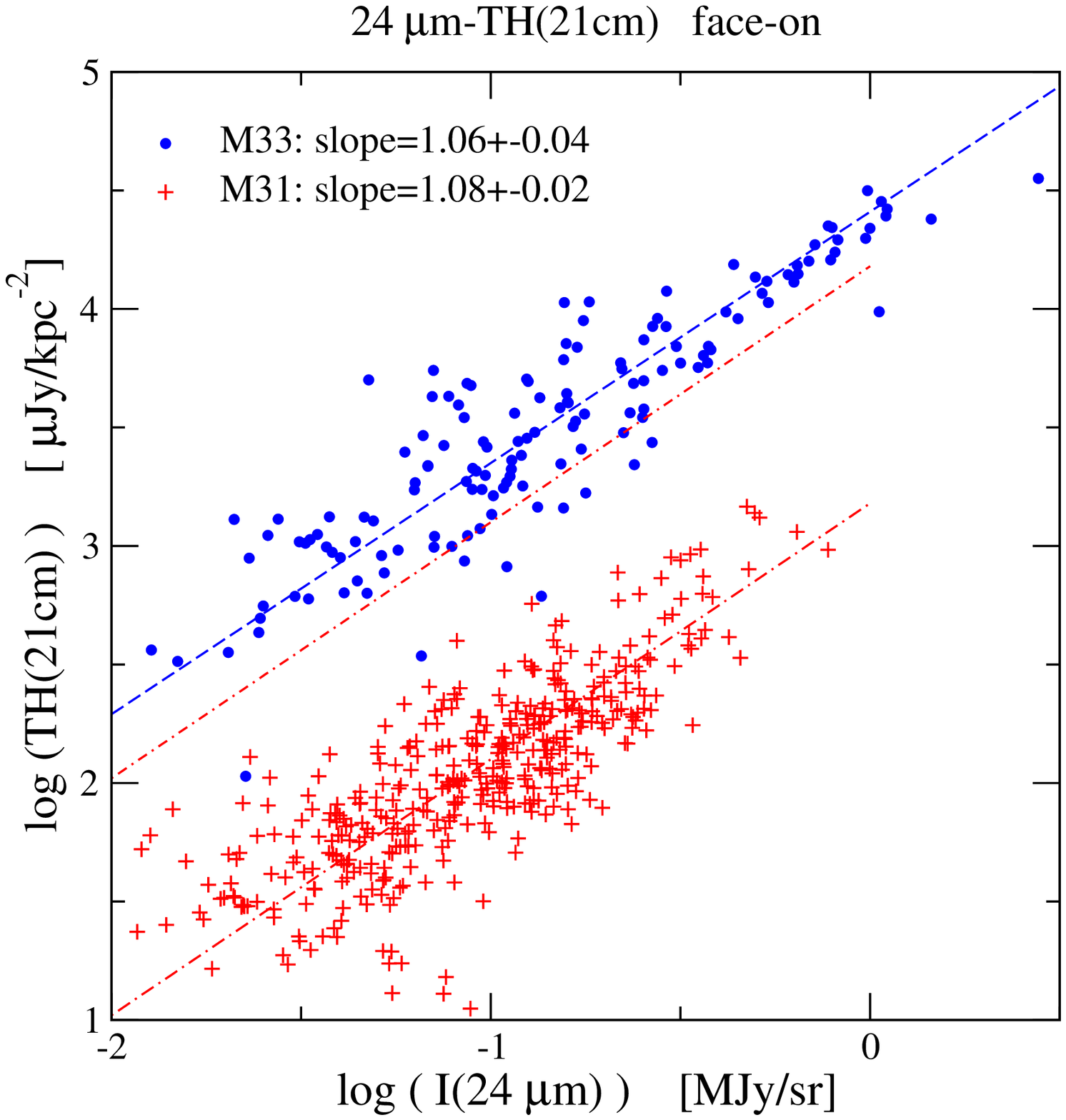}
\hspace{1.5cm}
\includegraphics[width=0.40\textwidth,angle=270]{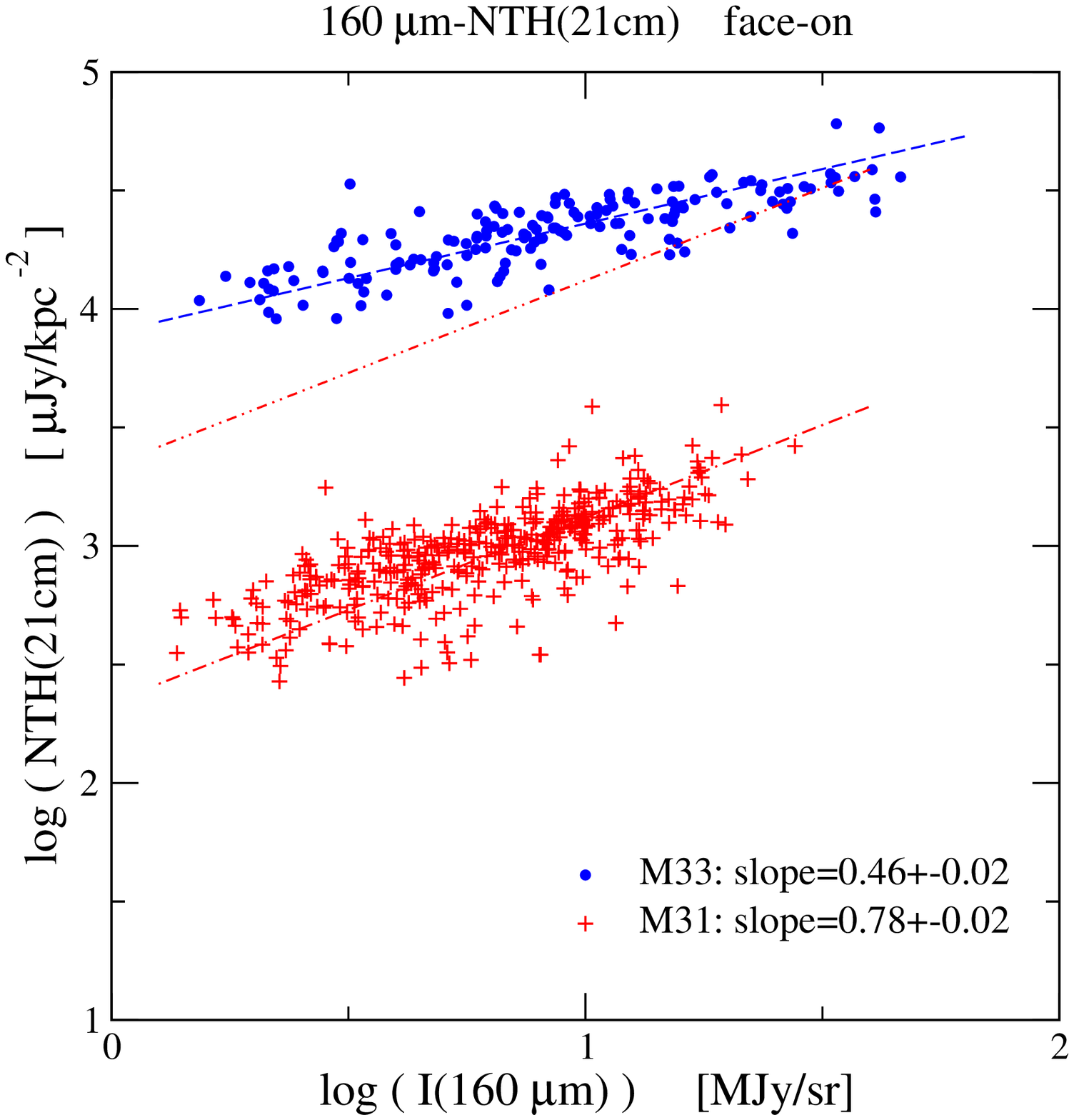}
\end{tabular}
\newline
\caption{Scatter plots between the surface brightnesses of FIR
emission and 21~cm radio continuum emission from M\,31 and M\,33.
Dashed lines show the power-law fits given in Tables~1 and 2. The
scatter plots for M\,31 are shifted down by a factor of 10 for
clarity. The correct locations of the fits are shown by the red
dashed lines just below the M\,33 plots. {\bf (a)} Thermal emission
at 21~cm versus $24\mum$ emission. {\bf (b)} Nonthermal emission at
21~cm versus $160\mum$ emission. A colour version of this
figure is shown in the electronic version of the paper.}
\label{fig1}
\end{figure*}

In Fig.~1 we present examples of scatter plots between radio and
dust emission for both galaxies. The correlations between thermal
radio emission at 21~cm and warm dust emission at $24\mum$ are close
to linear (Fig.~1a). This is not surprising as both emissions are
powered by massive stars. The power-law exponent of $b=1.08\pm0.02$
for M\,31 (Table~1) is consistent with the value of $1.2\pm0.1$
found by \cite{Hoernes98}. \cite{Xu92} derived a linear relationship
between thermal emission at 6.3~cm and warm dust emission for the
Large Magellanic Cloud.

Figure~1a shows that the thermal emission from M\,31 is weaker than
that from M\,33, which is due to its smaller star formation rate per
kpc$^2$ \citep{Tabatabaei10}. However, the same amount of thermal
emission corresponds to 70\% more $24\mum$ emission from M\,31 than
from M\,33. This indicates that a significant part of the $24\mum$
emission from the ``bright ring'' in M\,31 originates from sources
other than warm dust heated by massive stars. \cite{Tabatabaei10}
arrived at the same conclusion based on an excess in the $24\mum
/70\mum$ emission ratio.

For both galaxies the correlation between thermal emission and
emission from cool dust seen at $160\mum$ is non-linear with
exponent $b=1.4$ (Tables~1 and 2). Since the cool dust is mainly
heated by the diffuse interstellar radiation field (ISRF)
\citep{Xu92}, its emission is much smoother than
that of the ionized gas, leading to an exponent $b>1$.

Figure~1b shows that for both galaxies the correlation between
nonthermal emission at 21~cm and $160\mum$ emission has a power-law
exponent $b<1$, but the exponents for the two galaxies are
strikingly different. As the TH -- $160\mum$ correlations have the
same value of $b$, the difference is in the character of the
nonthermal emission that is much smoother in M\,33 ($b=0.46\pm0.02$)
than in M\,31 ($b=0.78\pm0.02$) caused by the relatively high
intensities towards lower brightness. This could mean that the
propagation length of the CRE in M\,33 is larger than in M\,31 or
that M\,33 has a halo.

The smaller values of exponent $b$ for M\,33 than for M\,31 are
confirmed by the correlations between nonthermal emission at 21~cm
and the emissions at $24\mum$ and $70\mum$ (Tables~1 and 2). In
Tables~1 and 2 we have also listed the power-law fits of the
correlations between nonthermal emission at 6.3~cm (M\,31) or
3.6~cm (M\,33) and dust emission. These have exponents much closer
to 1, but the difference between M\,31 and M\,33 remains. Exponents
closer to 1 imply less diffusion of CRE, which is expected
at smaller wavelengths. Below we will employ the observed
differences in the correlations to estimate the diffusion length of
the CRE in the two galaxies.

\section{Diffusion of CR electrons}

CR particles may propagate through the ISM by streaming with the Alfv{\'e}n
velocity along the ordered magnetic field lines or they may be scattered
by the irregularities of the turbulent magnetic field, leading to random-
walk diffusion along the ordered field. As in the Milky Way disk the
latter process dominates \citep {Strong07, Dogiel12}, we expect that
in the disks of M~31 and M~33 the CRE also mainly propagate by diffusion.

In order to estimate the diffusion length of CRE, we need to know
the total magnetic field strength, $\Btot$, in the galaxy disk,
which determines the energy and the lifetime of the CRE (see Eqs.~2
and 3 below). We derive $\Btot$ in Sect.~3.1 and the diffusion
lengths of CRE in the plane of the sky in Sect.~3.2.

\subsection{Synchrotron scale heights and magnetic field strengths}

We calculated the mean value of $\Btot$ in the radial intervals used
for the radio -- FIR correlations (i.e. $R=6.8-12.5\kpc$ in M\,31,
$R=0-5\kpc$ in M\,33) from the mean surface brightness of the
nonthermal (synchrotron) emission using the code BFIELD of M.~Krause
\footnote{www.mpifr-bonn.mpg.de/staff/mkrause} based on Eq.~3 of
\cite{Beck05}. This equation assumes energy equipartition between
total cosmic rays and total magnetic fields. The code asks for the
nonthermal spectral index $\alpha_\mathrm{n}$, the nonthermal
degree of polarization, the line of sight through the
medium $L$ and the ratio of the number densities of
protons and electrons $K$, here taken equal to 100. The value of
$\alpha_\mathrm{n}$ follows from the mean nonthermal intensities at
the two frequencies used ($I\propto \nu^{-\alpha_\mathrm{n}}$)
giving $\alpha_\mathrm{n}=0.92\pm0.06$ for M\,31 and
$\alpha_\mathrm{n}=0.86\pm0.08$ for M\,33 (see Table~6).
The degree of polarization, i.e. polarized/nonthermal intensity,
was calculated for the areas considered. The lines
of sight were obtained from the exponential scale heights of the
nonthermal emission estimated in the following way.

On the 21~cm nonthermal map of M\,31 at 45\arcsec\ resolution we
measured the half-power width of several spiral arms in each
quadrant of the bright emission ring along cuts parallel to the
minor axis taken at $X=25\farcm 5,\ 23\farcm 0$ and $19\farcm 5$
on the northern major axis and $X=14\farcm 5,\ 17\farcm 0$ and
$19\farcm 5$ on the southern major axis. Because of the high
inclination of M\,31, the cuts are nearly perpendicular to the arms
at these positions and the width of the arms on the sky plane is
determined by their vertical extent. We converted the half-power
widths, corrected for inclination and angular resolution, to
exponential scale heights and calculated the radius $R$ in the
galaxy plane for each measured position. This yielded mean scale
heights of $h=300\pm25\pc$ at $R=8-10\kpc$ and $h=335\pm15\pc$ at
$R=10-12\kpc$.

The mean exponential scale height in the radial interval
$R=6.8-12.5\kpc$, used for the radio -- FIR correlations, was
estimated from the scale height of the total gas (HI +
2H$_2$). \cite{Berkhuijsen93} noted that the scale height of the
total (mainly nonthermal) 21~cm emission is close to that of the
total gas in arms in the SW quadrant of the ``ring''. We checked
this on the total gas map of \cite{Nieten06}, smoothed to 45\arcsec\
resolution, by measuring the half-power widths of the gas arms at
the same positions as above. As we found a mean ratio of the
nonthermal/gas scale heights of $0.88\pm0.05$, the nonthermal scale
height at 21~cm is indeed nearly equal to the total gas scale
height. For the area $R=6.8-12.5\kpc$, the mean scale height of the
total gas is $h_\mathrm{gas}\simeq 330\pc$ (calculated from the
ratio mean column density / mean volume density
\citep{Tabatabaei10}, which is consistent with the nonthermal scale
heights derived above. Therefore, we adopted
$h_\mathrm{syn}=330\pm40\pc$ for M\,31 giving an effective line of
sight through the medium of
$L=2h_\mathrm{syn}/\cos(i)=2550\pm310\pc$.

For M\,33 we cannot measure the exponential scale height of the
21~cm nonthermal emission because at the inclination of 56\degr\ the
arm widths in and perpendicular to the disk are superimposed on the
sky plane. Therefore, we assumed the nonthermal scale height to be
equal to the scale height of the total gas, as is the case in
M\,31. Little information on the gas scale height in M\,33 exists in
the literature. We calculated the mean HI scale height for
$R<20\farcm 5$ from the half thickness as function of radius given
in Table~II of \cite{Warner73}, correcting for an erroneous factor
$\sqrt{2}$ and taking the increase in the area of rings with
increasing radius into account. For the distance $D=840\pc$ this
gave $h_\mathrm{HI}=360\pc$. \cite{Heyer04} found a relationship for
the column density ratio N(HI)/N(H$_2$) with radius. Integration out
to $R=20\farcm 5$ yielded a mean value of 3.4 giving
N(H$_2$)/N(HI)~$=0.29$. Assuming $h_\mathrm{H_2}=h_\mathrm{HI}/2$,
we obtained a scale height of the total gas of
$h_\mathrm{gas}=320\pc$. We adopted $h_\mathrm{syn}=320\pm80\pc$ and
$L=1140\pm280\pc$ for M\,33. Note that the uncertainty in $L$ has
little influence on the derived value of the magnetic field strength
because it enters the calculation to the power
$1/(3+\alpha_\mathrm{n})$.

Our estimates of $h_\mathrm{syn}$ agree with scale heights obtained
for other galaxies. Like in the Milky Way, observations of
edge-on galaxies usually reveal a thin synchrotron disk with a scale
height of about $0.3\kpc$ and a thick synchrotron disk with a scale
height of $1-2\kpc$ \citep{Beuermann85, Dumke98, Krause11}. Thus our estimates
of $h_\mathrm{syn} \simeq 330\pc$ refer to the thin disks in M\,31 and M\,33.

The derived scale heights and total magnetic field strengths are
summarized in Table~3 that also gives the scale heights of the CRE,
$h_\mathrm{CRE} = h_\mathrm{syn} (3+\alpha_\mathrm{n})/2$, and their
energy $E$ calculated from Eq.~3 in Sect.~3.2. The values of
$h_\mathrm{CRE}$ at the high frequencies were scaled from those at
1.4~GHz with $\nu^{-0.125}$ (see below Eq.~4 in Sect.~3.2).

\begin{table*}
\caption{Exponential scale heights and magnetic field strengths of the thin disks
in M\,31 and M\,33} \label{tab3}
\begin{tabular}{llcccccc}
\hline\hline Galaxy   &R    &Frequency &$h_\mathrm{syn}$
   &$h_\mathrm{CRE}\,\,^a$
   &$\Btot$  &$E$  &$t_\mathrm{syn}/2$ \\
   &$[\!\kpc]$  &$[\!\GHz]$ &$[\!\pc]$         &$[\!\pc]$
   &$[\!\muG]$ &$[\mathrm{GeV}]$  &$[10^7 \yr]$ \\
\hline
M31    &6.8-12.5   &1.465         &$330\pm40$\pheins  &$650\pm80$\pheins &$6.6\pm0.3$ &4.1  &$3.4\pm0.2$ \\
       &           &4.85\pheins   &--          &$560\pm70\,\,^b$     &$6.6\pm0.3$  &7.5  &$1.9\pm0.1$ \\
       &           &              &            &                   &             &     &  \\
M33    &0  - 5     &1.425         &$320\pm80\,\,^c$ &$620\pm150$\pheins &$8.1\pm0.5$ &3.7  &$2.5\pm0.2$ \\
       &           &8.35\pheins   &--             &$500\pm120\,\,^b$ &$8.1\pm0.5$  &8.9  &$1.1\pm0.1$ \\
\hline
\multicolumn{7}{l}{(a) $h_\mathrm{CRE} = h_\mathrm{syn} \, (3+\alpha_\mathrm{n})/2$ } \\
\multicolumn{7}{l}{(b) Scaled from $h_\mathrm{CRE}(1.4\GHz)$ with $\nu^{-0.125}$ (see Eq.~4) } \\
\multicolumn{7}{l}{(c) Assumed to be equal to the gas scale height like in M\,31 } \\
\end{tabular}
\end{table*}

\subsection{Diffusion lengths of CRE}

Since the sources of CRE probably are supernova remnants (SNRs) located
in or near star-forming regions, we used the distribution of the
{\em thermal}\ emission as the source distribution of the CRE
(see Fig.~2). We thus avoid the influence of dust properties
and other heating sources than massive stars (e.g. evolved stars,
the diffuse ISRF) that may play a role when the distribution
of IR emission is taken as the source distribution.
The relationships between nonthermal and thermal emission
(given at the end of Tables 1 and 2) show the same trends
as those between nonthermal and dust emission, i.e. exponents
$b$ increase with frequency, and for M\,33 the exponents are smaller
than for M\,31. We interpret these trends as the result of
differences in diffusion lengths of the CRE.

In the following we assume that CRE are injected into the ISM by SNRs
shocks continuously (at least during a few times the CRE lifetime),
leading to a stationary distribution of CRE in the galaxy. We
further assume that the diffusion of the CRE leads to a Gaussian
distribution of the CRE. The separation of the SNRs is larger than
the diffusion length of the CRE in the disk and avoids overlap of
diffusion lengths of CRE from neighbouring sources, which would
otherwise influence the estimated mean diffusion length. Within a
radius of 5\,kpc from the centre, M\,33 contains $\lesssim 40$ SNRs and
SNR candidates of diameter $D\lesssim 30$\,pc \citep{Long10}, which are
the main sources of CRE \citep{Berezhko04}. Their mean separation is
1.4\,kpc, which is larger than the typical diffusion length of
$\simeq$ 1\,kpc. As the star formation rate / area in the emission ring
of M\,31 is five times smaller than in M\,33 (see Table~6), the mean
separation of CRE sources in the ring is \,$> 1.4$\,kpc. So it will be
possible to estimate the CRE diffusion length in both galaxies.

The only difference between the nonthermal emissions from a galaxy
at two frequencies is the diffusion length of CRE that is reached by
random-walk diffusion, other variables like their source
distribution and the magnetic field strength being the same.
The diffusion length $l(\nu)$ at frequency $\nu$ is determined by the
synchrotron life time $t_\mathrm{syn}$ and the diffusion coefficient
\footnote{$1\pc^2 \yr^{-1} \simeq 30.2\,10^{28}\cm^2\, \mathrm{sec}^{-1}$}
$D_\mathrm{E}$. The mean synchrotron lifetime  \,$<t>$\, of the bulk of
CRE, a mixture of CRE of all ages, is smaller than $t_\mathrm{syn}$; it
may be written as
\begin{equation}
<t> = \int\limits_{0}^{t_\mathrm{syn}} N \, t \, dt \, / \, \int\limits_{0}^{t_\mathrm{syn}} N \,dt,
\end{equation}
where $N$ is the number of CRE as a function of time,
$N = N_\mathrm{0} \, \exp[-t/t_\mathrm{syn}]$. Evaluation of \,$<t>$\, yields
\,$<t> \,= t_\mathrm{syn} (1-2/e) \, / \, (1-1/e) = 0.42 \, t_\mathrm{syn}$.
Therefore, we take $t_\mathrm{syn}/2$ as the typical age of the
bulk of the CRE in the calculations below. The diffusion length then is
\begin{equation}
(l_\nu/\mathrm{pc})^2 \,\, \simeq \,\,
(D_{\mathrm{E},\nu}/\mathrm{pc}^2 \yr^{-1}) \,\,
(t_\mathrm{syn}/2)/10^7\yr.
\end{equation}
$D_{\mathrm{E},\nu}$ depends on the energy $E$ of the CRE
as $D_{\mathrm{E},\nu}\propto E^p$, where $p\approx0.5$
\citep{Strong07}. $E$ is related to the radiating frequency:
\begin{equation}
(\nu /\mathrm{GHz}) \,\, \simeq \,\, 1.3 \, 10^{-2} \,\, (\Btot /\mu
\mathrm{G}) \, \, (E/\mathrm{GeV})^2,
\end{equation}
where $\Btot$ is the total magnetic field strength derived from
the surface brightness of the nonthermal emission. Mean values of
$\Btot$ and $E$ for the regions used for the correlations are
given in Table~3.

The lifetime of the radiating CRE is mainly determined by the
synchrotron losses at our frequencies (e.g. Fig.~1 in
\cite{Murphy09}):
\begin{equation}
(t_\mathrm{syn}/10^7\yr) \,\, \simeq \,\, 1.4\,10^9  \,\,
(\nu/\!\mathrm{GHz})^{-0.5} \, \, (\Btot/\!\muG)^{-1.5}.
\end{equation}

Thus, at high $\nu$ we see radiation from younger electrons than
at low $\nu$, and because of their shorter lifetime they cannot travel
as far as those at low $\nu$, in spite of their higher energy.
Comparing Eqs.~1, 2, and 3, we see that for a fixed value of
$\Btot$, $\, l_\nu^{~2} \propto \nu^{-0.25}$, if
$D_{\mathrm{E},\nu}\propto E^{0.5}$. An estimate of
$l_\nu^{~2}$ yields an estimate of $D_{\mathrm{E},\nu}$ from Eq.~2.

We measured the diffusion length in the plane of the sky in two
ways: (1) by determining the difference in diffusion length between
1.4~GHz and the higher frequency, and (2) by estimating the diffusion
length at each frequency separately using the distribution of the
thermal emission as the source distribution of the CRE.

1. The most direct way to measure the average diffusion length
of CRE is a comparison of the nonthermal distributions at the two
frequencies. Since the CRE radiating at the higher frequency
are still closer to their origin than at 1.4~GHz, smoothing of the
{\em nonthermal}\ distribution at the higher frequency will make it
more similar to that at 1.4~GHz. We smoothed the high-frequency
distributions until the power-law exponent of the correlation
with the nonthermal distribution at 1.4~GHz had  $b=1$, indicating
that the two distributions are most similar. The smoothing beam required
for this, corrected for the resolution of the observations, represents the
difference in diffusion length between the two frequencies.

Because distances are measured from the centre of the telescope beam,
we took half of the full width at half power of the corrected smoothing beam
as the derived difference in the square of the diffusion lengths,
$\Delta~l^2$. Then, using
Eq.~1 and $\Delta\,l^2 = l_{1.4}^{~2} - l_\nu^{~2}$, we have
\begin{equation}
\Delta\,l^2 \,\, \simeq \,\, l_{1.4}^{~2} \,\, (1 -
(\nu/1.4\,\mathrm{GHz})^{-0.25}),
\end{equation}
where $l_{1.4}$ is the diffusion length at 1.4\,GHz. From the
derived value of $l_{1.4\GHz}^{~2}$ we obtain $l_\nu^{~2}$ by
scaling with $\nu^{-0.25}$. The resulting diffusion lengths observed
in the plane of the sky, $l_\mathrm{sky}$, are given in Table~4. The
errors are determined by the standard deviations in $b$ of the
correlations giving the best smoothing beam.

\begin{table}
\caption{Observed diffusion lengths of CRE in M\,31 and M\,33 on the sky}
\label{tab4}
\begin{tabular}{l l l l}
\hline\hline
Galaxy   &Method   &Frequency   &$l_\mathrm{sky}$ \\
         &         &$[\!\GHz]$  &$[\!\pc]$        \\
\hline
M31      &1        &1.465       &$1140\pm 190\,\,^a$        \\
         &         &4.85        &\pheins $980\pm 160\,\,^b$ \\
         & \\
         &2        &1.465       &\pheins $715\pm ~20$      \\
         &         &4.85        &\pheins $510\pm ~60$      \\
\hline
M33      &1        &1.425       &\pheins $900\pm 110\,\,^a$ \\
         &         &8.35        &\pheins $720\pm ~90\,\,^b$  \\
         & \\

         &2        &1.425       &$1960\pm 170$            \\
         &         &8.35        &$1450\pm 200$            \\
\hline \multicolumn{4}{l}{(a) Based on the observed difference in
diffusion length between two}\\
\multicolumn{4}{l}{frequencies (see Eq.~4):}\\
\multicolumn{4}{l}{M\,31: $(\Delta\,l^2)^{0.5} = 580\pm
100\pc$; M\,33: $(\Delta\,l^2)^{0.5} = 540\pm 65\pc$ }\\
\multicolumn{4}{l}{(b) Scaled from 1.4~GHz with
$l_\mathrm{sky}\propto \nu^{-0.125}$ }
\end{tabular}
\end{table}

2. The exponents $b$ of the nonthermal--thermal correlations are
$<1$ because the CRE diffused away from their places of origin
represented by the distribution of the thermal emission ( see
Sect.~4 and Fig.~2). We obtained an estimate of the diffusion length
at each frequency by smoothing the {\em thermal}\ emission until the
exponent of the correlation with the nonthermal emission at that
frequency became $b=1$. By requiring $b=1$, we implicitly assume that
$\Btot^{~1+\alpha_\mathrm{n}}$ is about constant within the area
considered ($I_\mathrm{NTH} \propto N_\mathrm{CRE} \,
\Btot^{~1+\alpha_\mathrm{n}}$). On large scales this is
approximately the case as radial variations in $\Btot$ are $<10\%$
in the areas used for the correlations (i.e. see Fig.~9 in
\cite{Tabatabaei08}). Small-scale variations in $\Btot$ will
only contribute to the spread in the data points. Again, we took
half the corrected smoothing beam as the diffusion length in the sky
plane to obtain the diffusion lengths given in Table~4.

In both cases we used circular Gaussian smoothing beams because they
gave the best fits. Even for the highly inclined galaxy M\,31, fits
with elliptical beams parallel to the major axis were worse than
those with circular beams. This agrees with the results of
\cite{Marsh98} who found circular smoothing beams optimal for 10 out
of 16 galaxies. Hence, in these galaxies the deprojected diffusion
lengths within the disk and perpendicular to the disk are about
the same (see Sect. 5).

\section{Diffusion lengths in the plane of the sky}

The observed diffusion lengths in Table~4 are in the range
$0.5-2\kpc$ in the plane of the sky (measured along the major axis).
At each frequency, the two methods yield diffusion lengths that
agree within a factor of 2. Method 2 yields smaller values than
method 1 for M\,31, but larger values for M\,33. The estimates
by the two methods may differ for several reasons.

a) Method 1 measures the difference in diffusion length between the
two frequencies. Since the lifetime of the CRE at the higher
frequency is about half of those at 1.4\,GHz (see Table~3), they do
not propagate as far as those at 1.4\,GHz. This means that the
diffusion length of the older CRE emitting at 1.4\,GHz, which had
time to propagate far from their places of origin, determines
$\Delta\,l^2$. Hence, $\Delta\,l^2$ is an upper limit to the mean
diffusion length of the bulk of CRE. If older CRE not only propagate
along the disk but also move out of the thin disk into a thick
disk / halo, the ``vertical'' propgation length may add to the
diffusion length observed in the thin disk by an amount depending on
the inclination. However, observations of edge-on galaxies show that
CRE propagation out of the thin disk usually occurs in SF regions
(i.e. \cite{Dahlem94, Dumke95, Heesen09}) and most of these CRE will
be young \citep{Breitschwerdt02}. Therefore we expect that method 1
estimates an upper limit to the diffusion length at 1.4\,GHz that
mainly refers to the thin disk.

b) Method 2, however, measures the mean diffusion length of the bulk
of the CRE, i.e. of a mixture of CRE of all ages, at each frequency.
The inclusion of young CRE will yield a diffusion length smaller than
the upper limit from method 1, consistent with our results for M\,31.
Furthermore, method 2 includes the full effect of CRE propagation
out of the thin disk. If this is significant, the observed diffusion length
could be larger than the estimate from method 1, as we find for M\,33. We
discuss the evidence for a thick disk / halo in M\,33 in Sect.~5.2.

c) Method 2 is based on the assumption that the source distribution of
the CRE is similar to that of the thermal emission. Differences between
the results of methods 1 and 2 could occur if this assumption is incorrect.
However, Figure~1 of \cite{Long10} shows that most of the 137 known
supernova remnants and SNR candidates in M\,33 are located on or close
to H$\alpha$ emission. We used these data to construct the radial
distribution of the number of SNRs/kpc$^2$ in M\,33 presented in Figure~2,
which also shows the radial variation of the thermal emission. Out to
$R=5\kpc$, the radial distribution of the SNRs is very similar in
shape to that of the thermal emission and has roughly the same scale
length. As in the dense ISM near the centre not all SNRs may have
been detected, the scale length of $2.9\pm0.7\kpc$ may be an upper
limit. Thus in M\,33 the distribution of the thermal emission may
indeed be representative for the source distribution of CRE.

Recently, \cite{Tabatabaei13b} estimated the
diffusion length of CRE on small scales ($<1\kpc$) in M\,31 and
M\,33 from NTH--FIR correlations as a function of scale.
\footnote{Using a wavelet analysis, \cite{Tabatabaei13b}
decomposed each IR and radio map into a set of 6 (M\,31) or 7 (M\,33) maps
each containing emission on a specific scale; the scales chosen are
between 0.4 and 4\,\,kpc. For each galaxy, maps on the same scale are
are then correlated and the derived correlation coefficients ($r_w$)
are plotted as a function of scale. The smallest scale on which the
synchrotron -- $70\,\mum$ correlation has $r_w=0.5$ is taken as the
diffusion length of the CRE, because $r_w < 0.5$ implies that the
correlation is lost, i.e.\, due to CRE diffusion and/or escape and
turbulence in the ISM caused by SF regions and SNRs, synchrotron
and $70\,\mum$ emission do not correlate on scales of $\lesssim 1$\,kpc.}
They find $l_\mathrm{sky}=730\pm90\pc$ for M\,31 and $l_\mathrm{sky}<400\pc$
for M\,33. Their value for M\,31 is in excellent agreement with
$l_\mathrm{sky}=715\pm20\pc$ that we obtain from method 2, but their
value for M\,33 is lower than our upper limit of
$l_\mathrm{sky}=900\pm110\pc$ for the thin disk. As the authors
explain, this low $l_\mathrm{sky}$ for M\,33 refers mostly to young
CRE that are still close to their places of origin in the star
forming regions. So their value may be considered a lower limit to
$l_\mathrm{sky}$ of the bulk of CRE in M\,33. Since the emission
from star forming regions occurs on small scales, their value is free
from extended thick disk / halo emission.

\begin{figure}
\centering
\includegraphics[width=0.7\columnwidth,angle=270]{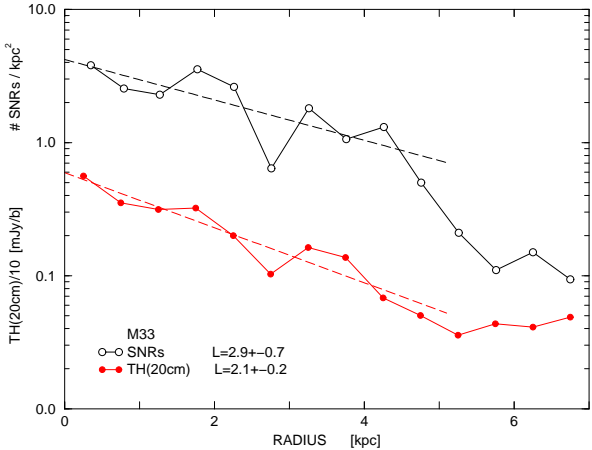}
\caption{Radial distributions in M\,33: {\it black} -- surface
density of SNRs and SNR candidates of \citet{Long10}; {\it red} --
surface brightness of thermal emission at 21~cm. Dashed lines are
fits to the function $exp(-R/L)$ with scale length $L$ (in kpc)
given in the figure. The similarity between the two distributions
indicates that the the distribution of thermal emission is representative
for that of the SNRs, the source distribution of the CRE. }
\label{fig2}
\end{figure}


\section{Distribution of CRE in M\,31 and M\,33}
\label{cre}

Table~4 shows that at 1.4~GHz the diffusion length in the sky plane
is about $1\kpc$ for M\,31 and about $1.5\kpc$ for M\,33. We have
deprojected $l_\mathrm{sky}$ into the components $l_{x,y}$ in the
plane of the galaxy and $l_z$ perpendicular to the disk by assuming
istropy in the disk of the galaxy. Isotropy of CRE  is consistent with
the situation in the Milky Way where the diffusion of cosmic rays becomes
isotropic on scales larger than 100--200\,pc \citep{Strong07, Dogiel12}, but
below 1\,kpc \citep{Stepanov12}. Isotropy of CRE on large scales is also
expected from simulations \citep{Yan08}.

We thus have $l_{sky} = l_x = l_y = l_z$, where $l_x$ and $l_y$ are
measured along the major and minor axis in the plane of the
galaxy, respectively, giving $l_{x,y} = l_x \sqrt{2}$ and $l_z =
l_x$. This assumption leads to a circular extent of the propagation
lengths on the sky plane, in agreement with the circular smoothing
beams we derived. We present these values in Table~5.

In Table~5 we also give diffusion coefficients and confinement times
of the CRE in the two galaxies. The diffusion coefficients in the
x, y and z directions, $D_\mathrm{Ex,y}$ and $D_\mathrm{Ez}$, follow
from Eq.~2 with the typical CRE ages $t_\mathrm{syn}/2$ listed in
Table~3. The confinement time is the time the CRE need to reach the
scale height of the thin disk by diffusion, $t_\mathrm{conf} =
h_\mathrm{CRE}^2 / D_\mathrm{Ez}$. We also list the mean velocities
of the CRE when crossing the thin disk, $V_\mathrm{CRE} =
h_\mathrm{CRE} / t_\mathrm{conf}$. We discuss these results below.
Note that since the diffusion lengths estimated by method 1 are
upper limits, the values for the  diffusion coefficients also are upper
limits,
\footnote{Since method 1 refers to the older CRE, we should rather use
$3 t_\mathrm{syn}/4$ than $t_\mathrm{syn}/2$ to derive the diffusion
coefficients, which would lower the upper limits by 30\%. But in view
of the large errors and for simplicity, we use $t_\mathrm{syn}/2$.}
 those for $t_\mathrm{conf}$ lower limits and those for
$V_\mathrm{CRE}$ upper limits.

\begin{table*}
\caption{Diffusion lengths and confinement times of CRE in M\,31 and M\,33}
\label{tab5}
\begin{tabular}{l l c c c c c c c}
\hline\hline GAL   &M  &FREQ  &$l_\mathrm{x,y}\,\,^a$
&$D_\mathrm{Ex,y}$ &$l_\mathrm{z}\,\,^b$  &$D_\mathrm{Ez}$
   &$t_\mathrm{conf}\,\,^c$  &$V_\mathrm{CRE}\,\,^d$ \\
      &   &$[\!\GHz]$  &$[\!\pc]$  &$[10^{28}\cmsqs]$  &$[\!\pc]$
   &$[10^{28}\cmsqs]$  &$[10^7\yr]$  &$[\!\kms ]$  \\
\hline
M31  &1  &1.465  &$1610\pm280$  &$2.3\pm0.8$  &$1140\pm200$
   &$1.2\pm0.4$  &$1.1\pm0.5$  &$\pheins 60\pm \pheins 25$ \\
     &   &4.85\pheins   &$1390\pm230$  &$3.1\pm1.1$  &$\pheins  980\pm160$
   &$1.5\pm0.5$  &$0.6\pm0.3$  &$\pheins 90\pm \pheins 40$  \\
     &  &  &  &  &  &  &  &  \\
     &2  &1.465  &$1010\pm \pheins 30$  &$0.9\pm0.1$  &$\pheins 715\pm \pheins 20$
   &$0.45\pm0.03$  &$2.8\pm0.7$  &$\phelf 20 \pm \phelf 6$ \\
     &   &4.85\pheins  &$720\pm \pheins 90$  &$0.8\pm0.2$  &$\pheins 510\pm \pheins 60$
   &$0.41\pm0.11$  &$2.3\pm0.9$  &$\phelf 25 \pm \phelf 8$ \\
\hline
M33  &1  &1.425  &$\pheins 1270\pm160$  &$1.9\pm0.6$  &$\pheins 900\pm110$
   &$1.0\pm0.3$  &$1.2\pm0.6$  &$\pheins 50 \pm \pheins 30$ \\
     &   &8.35\pheins  &$\pheins 1020\pm130$  &$2.9\pm0.8$ &$\pheins 720\pm \pheins 90$
   &$1.4\pm0.4$  &$0.5\pm0.3$  &$\pheins 100\pm \pheins 55$ \\
     &  &  &  &  &  &  &  &  \\
     &2  &1.425  &$2770\pm240$  &$9.3\pm1.9$  &$1960\pm170$
   &$4.6\pm0.9$  &$0.3\pm0.2$  &$210\pm120$  \\
     &  &8.35\pheins   &$2050\pm290$  &$11.5\pm3.5$  &$1450\pm200$
   &$5.8\pm1.8$  &$0.13\pm0.08$  &$380\pm230$ \\
\hline
\multicolumn{9}{l}{(a) $l_\mathrm{x,y} = l_\mathrm{sky} \sqrt{2}$ } \\
\multicolumn{9}{l}{(b) $l_\mathrm{z} = l_\mathrm{sky}$ } \\
\multicolumn{9}{l}{(c) $t_\mathrm{conf} = h^2_\mathrm{CRE} /
D_\mathrm{Ez}$
(with $h_\mathrm{CRE}$ from Table~3) } \\
\multicolumn{9}{l}{(d) $V_\mathrm{CRE} = h_\mathrm{CRE}/t_\mathrm{conf}$ } \\
\end{tabular}
\end{table*}

\subsection{Confinement of CRE in M\,31}

The diffusion length $l_\mathrm{x,y}$ in the plane of M\,31 of about
1-2~kpc at 1.4~GHz is typical for galaxy disks \citep{Murphy06b}.
The corresponding diffusion coefficient of about
$(1-3)\,10^{28}\cmsqs$ is in the range $10^{28}-10^{29}\cmsqs$
estimated for the Milky Way and other galaxies (e.g. \cite{Dahlem94,
Ptuskin98, Moskalenko02, Murphy09, Heesen09, Lacki10}, and
references therein; \cite{Buffie12, Tabatabaei13a}) .

Perpendicular to the disk the diffusion length at 1.4~GHz  is only
about $0.7-1.1\kpc$. Since method 1 yields upper limits for the
diffusion lengths, we use the smaller value of $l_\mathrm{z}$ of
method 2 for comparison with the exponential scale height
$h_\mathrm{CRE}$ given in Table~3. We convert the Gaussian-based
value of $l_\mathrm{z}$ at 1.4~GHz to the exponential scale height
\footnote{As for a normalized Gaussian the halfwidth $l$ occurs at
z=0.833 and for a normalized exponential the scale height $h$ is
reached at $z=1$, $h=l/0.833$.} $h_\mathrm{z} = l_\mathrm{z}/0.833 =
860\pm25\pc$, which is about 30 percent higher than the value of
$h_\mathrm{CRE} = 650\pm80\pc$ given in Table~3. Both values refer
to a mixture of CRE of all ages. Since $h_\mathrm{CRE}$ is derived
from $h_\mathrm{syn}$ assuming energy equipartition between magnetic
fields and CR particles, the fair agreement with $h_\mathrm{z}$
suggests that energy equipartition may indeed be valid for the
radial range $R=6.8-12.5\kpc$ in M\,31.

The corresponding diffusion coefficient $D_\mathrm{Ez}$ of $0.45\,
10^{28}\cmsqs$ is about half of  $D_\mathrm{Ex,y}$ and very low,
indicating that CRE only slowly diffuse away from the midplane of
the disk. The time the CRE need to move out of the thin disk,
$t_\mathrm{conf} = (2.8\pm0.7)\, 10^7\yr$, is close to the typical
CRE age $t_\mathrm{syn}/2 = (3.4\pm0.2)\, 10^7\yr$ (see Table~3).
Hence, the CRE have lost most of their energy before they can reach
$z = h_\mathrm{CRE}$ and only older CRE may move into a thick
disk / halo, as the lower limit of $t_\mathrm{conf}$ of method 1
suggests. Thus, while the bulk of the CRE in M\,31 is confined to the
thin disk (as shown by method 2), some of the older CRE may leave
the thin disk at speeds of $V_\mathrm{CRE} < 60\pm25\kms$. As this
is about equal to the Alfv{\'e}n speed \footnote{We derived the
Alfv{\'e}n speed $V_\mathrm{A} (\pc/10^{-6} \yr) = B_\mathrm{perp,
ord} (\muG) / (4 \pi <n_\mathrm{e}> (\cm^{-3})^{-0.5})$, where
$<n_\mathrm{e}>$ is the density of free electrons in the ISM
obtained from the mean rotation measure \citep{Fletcher04} and
$B_\mathrm{perp,ord}$ is the strength of the ordered magnetic field
perpendicular to the LOS calculated from the polarized synchrotron
emission.} of $V_\mathrm{A}=50\pm20\kms$, convective propagation of
CRE by gas flows from SF regions does not play a role in M\,31.

Also \cite{Graeve81} and \cite{Berkhuijsen91} concluded that
M\,31 does not have a thick disk / halo. The confinement of the CRE to
the thin disk in M\,31 is consistent with the structure of the ordered
magnetic field. \cite{Fletcher04} found no significant vertical
component of this field that is extraordinarily well aligned along the
spiral arms in M\,31. The strong uniformity of the ordered field in
the disk \citep{Berkhuijsen03} probably leads to the small value of
$D_{Ez}$ and the confinement of the CRE to the thin disk.

\subsection{Distribution of CRE in M\,33}

An analysis of the polarized emission from M\,33
\citep{Tabatabaei08} revealed that the large-scale, ordered magnetic
field in M\,33 has a vertical component of about $1\muG$, extending to
at least $R=5\kpc$. This implies that M\,33 has a nonthermal halo or thick
disk containing CRE. In spite of this, the upper limits to the
diffusion lengths obtained by method 1 (see Table~5) are smaller
than the estimates of method 2, and even smaller than those of method 1
for M\,31. It seems that the halo contributions to the diffusion lengths
at the two frequencies largely canceled in the difference $\Delta\,l^2$
(see Equation~5). We conclude that the results of method 1 apparently refer
to the thin disk in M\,33. As the estimates of method 2 refer to thin+thick
disk (Sect.~4), we discuss these cases separately.

THIN DISK. The upper limit to the diffusion length $l_\mathrm{x,y}$
of 1.3~kpc at 1.4~GHz obtained from method 1 for the thin disk of M\,33 is
about 30 percent smaller than that in M\,31 (see Table~5). As the stronger
magnetic field $\Btot$ (Table~3) yields smaller synchrotron loss times, the
diffusion coefficient $D_\mathrm{Ex,y}$ of $(1.9\pm0.6)\, 10^{28}\cmsqs$ is
similar to that in M\,31.

The upper limit to the diffusion length perpendicular to the disk
of $l_\mathrm{z}=900\pm110\pc$ and $D_\mathrm{Ez}$ are also similar within
errors to those in M\,31 at 1.4~GHz. Here $l_\mathrm{z}$ corresponds to an
exponential scale height $h_z=l_\mathrm{z}/0.833=1080\pm130\pc$, an
upper limit consistent with the scale height $h_\mathrm{CRE}=620\pm150\pc$
listed in Table~3. As the lower limit to the confinement time is about half
of $t_\mathrm{syn}/2$, some of the older CRE traced by method 1 may leave
the thin disk at velocities of $V_\mathrm{CRE}= 50\pm30\kms$.
Like in M~31, this is equal to the Alfv{\'e}n speed
$V_\mathrm{A}=45\pm25\kms$.

THIN+THICK DISK. At 1.4~GHz the diffusion length $l_\mathrm{z}$ of
about 2~kpc, more than twice that of the thin disk, yields an
exponential scale height of $h_\mathrm{z}=2350\pm280\pc$,
corresponding to a synchrotron scale height of
$h_\mathrm{syn} = 2h_\mathrm{z}/(3+\alpha_\mathrm{n})=1220\pm270\pc$.
This value is consistent with the scale heights near 1~kpc in total emission
observed in edge-on galaxies \citep{Hummel91, Dumke95}, and  confirms the
existence of a thick synchrotron disk or halo in M\,33 \citep{Tabatabaei08}.
We may estimate the exponential scale height of the thick synchrotron disk
by subtracting the contribution of the thin disk to $h_\mathrm{syn}$, assuming
that at z=0~kpc the intensities of thin and thick disk are equal. With
$h_\mathrm{syn}=1.22\pm0.27\kpc$ and $h_\mathrm{thin}=0.32\pm0.08\kpc$
(Table~3), we find $h_\mathrm{thick}=2.1\pm0.5\kpc$ at 1.4~GHz. This
value of $h_\mathrm{thick}$ agrees with the mean value of
$h_\mathrm{thick}=1.8~\kpc$  in 5 galaxies (at 4.8~GHz) listed by
\cite{Krause11}. The agreement, however, may be fortuitious because
(1) our estimate of $h_\mathrm{thick}$ in M33 is sensitive to the
relative intensities of thin and thick disk at z=0~pc, and (2) the
value of Krause (2011) refers to the sum of thermal and nonthermal
emission instead of to nonthermal emission alone.

The value of $D_{Ez}$ of $4.6\, 10^{28}\cmsqs$ is nearly 5 times higher
than in the thin disk. This leads to a confinement time of
$t_\mathrm{conf} =(0.3\pm0.2)\, 10^7\yr$ that is significantly
shorter than the typical CRE age $t_\mathrm{syn}/2 = (2.5\pm0.2)\,
10^7\yr$. Hence, in M\,33 many CRE will move out of the thin disk
into the thick disk / halo at a mean velocity of $V_\mathrm{CRE}=
210\pm120\kms$. This velocity is higher than the Alfv{\'e}n velocity
of about $50\kms$, suggesting that CRE are leaving the disk by a
galactic wind.

As the older CRE traced by method 1 are largely confined to
the thin disk, mainly young CRE are moving along the vertical
magnetic field into the thick disk / halo. The multi-scale analysis of
the NTH--FIR correlation in M\,33 of \cite{Tabatabaei13b}
indicates that CRE propagate from the star
forming regions into the thick disk. Outflow of CRE and gas from
bright SF regions is often observed in edge-on galaxies with a thick
radio disk or halo (e.g. \cite{Dahlem95, Krause09, Heesen09}). The
outflow weakens the NTH emission from SF regions and decreases the
dynamic range of the emission from the thin disk. CRE in the thick
disk / halo also have velocity components parallel to
the disk, leading to distribution of the CRE across the disk and
further smoothing of the observed NTH emission
\citep{Breitschwerdt02, Dogiel12}. Hence, outflow of CRE from SF
regions into the thick disk / halo mimics the effect of ``horizontal''
CRE diffusion in the thin disk.

The large value of $l_\mathrm{x,y}\simeq 2.8\kpc$ obtained by method 2
at 1.4~GHz, more than twice as large as found for the thin disk, is
consistent with the effects of CRE outflow described above. The
smoothness of the observed NTH emission from the (thin + thick) disk
of M\,33 is reflected in the large $l_\mathrm{sky}$ of about 2~kpc
(Table~4) and in the low exponents $b$ of the NTH--FIR correlations
(Table~2).

\section{Discussion}

In this section we discuss the effect of the superposition of the
disk and halo emission in M\,33 along the LOS as well as the influence
of the magnetic field structure and of the star formation rate on the
CRE distribution in M\,31 and M\,33.

\subsection{Effect of the superposition on the sky of disk and halo emission }

In Section~5.2 we have argued that M\,33 has a nonthermal thick
disk or halo. Since the inclination angle of M\,33 is 56\degr, this
means that on the sky plane we see the halo emission superimposed
onto the more prominent disk emission. The addition of the
distributions of two components generally leads to a smoother
distribution than that of the individual ones. Therefore, the
nonthermal emission observed on the sky will be smoother than that
of the disk emission alone. This implies that the derived diffusion
length will be larger than the actual diffusion length in the thin
disk, because (using method 2) the TH emission needs to be more
strongly smoothed to reach exponent $b$=1 in $NTH\propto TH^b$. This
explains why in M\,33 $l_\mathrm{x,y}$ obtained by method 2 (Table~5)
is larger than that of the thin disk found by method 1 although the
CRE traced by method 2 are younger than those traced by method 1.

Possibly, not only $l_\mathrm{x,y}$ but also $l_\mathrm{z}$ is larger
than the actual propagation length perpendicular to the disk. In that
case also the derived scale heights will be too high. Since we cannot
separate the disk and halo emission from M\,33, the values of method 2
given in Table~5 should be regarded as upper limits.

We conclude that for galaxies with a thick NTH disk or halo the
apparent diffusion length estimated from a comparison with H$\alpha$
or FIR data may be considerably larger than the actual diffusion length
of the CRE in the thin disk, and that also the extent of the thick disk / halo
may be overestimated.

\subsection{Importance of the magnetic field structure}
\label{structure}

The diffusion length $l_\mathrm{x,y}$ in the thin disk of M\,31 at
1.4~GHz is about 1.3 times that in M\,33 (Table~5, method 1).
This difference may be due to differences in the diffusion coefficient
and synchrotron lifetime since $l=(D_\mathrm{E} \, t_\mathrm{syn}/2)^{0.5}$
(Eq.~2).

The propagation of CRE in the ISM is largely determined by their
interaction with magnetic fields. The diffusion coefficient depends
on both the turbulent and the ordered magnetic field strength. The
ordered field, mainly running parallel to the disk, favours the
propagation of CRE along the disk, while the turbulent field hinders
the CRE in traveling long distances in one direction because they
are scattered by the fluctuating magnetic field (MHD waves). Models
of diffusive CR propagation indicate that $D_E \propto
(\Btur/\Bord)^{-2}$ \citep{Chuvilgin93, Breitschwerdt02, Yan04,
Yan08, Shalchi10, Dogiel12}. Table~6 gives $\Btur$, $\Bord$ and
their ratio in the thin disks of M\,31 and M\,33. The higher $\Btur$
and lower $\Bord$ in M\,33 yield a $2.3\pm0.8$ times higher value of
$\Btur/\Bord$ and one would expect a $5.3\pm2.6$ times lower value
of $D_\mathrm{Ex,y}$ in M\,33 than in M\,31. However, the observed
factor obtained from Table~5 is only $1.2\pm0.6$, which suggests
that other factors also play a role.

\begin{table}
\caption{Mean values of some relevant variables in M\,31 and M\,33 }
\label{tab6}
\begin{tabular}{l l c c}
\hline\hline
Variable &Unit   &M\,31   &M\,33 \\
\hline
R                     &$\!\kpc$          &$6.8-12.5$    &$0-5$  \\
NTH(21~cm)            &$\!\muJy/\kpc^2$  &$7.3\pm0.4$   &$17.3\pm0.8$ \\
$\alpha_\mathrm{n}$   &--                &$0.92\pm0.06$ &$0.86\pm0.08$ \\
$\Btot$               &$\!\muG$          &$6.6\pm0.3$   &$8.1\pm0.5$ \\
$\Btur$               &$\!\muG$          &$5.0\pm0.2$   &$7.6\pm0.5$ \\
$\Bord$               &$\!\muG$          &$4.3\pm0.3$   &$2.8\pm0.3$ \\
$\Btur /\Bord$        &--                &$1.2\pm0.4$   &$2.7\pm0.3$ \\
TH(21~cm)             &$\!\muJy/\kpc^2$  &$1.1\pm0.1$   &$5.35\pm0.95$ \\
$f_\mathrm{th}$(21~cm)&$\!\%$            &$13\pm1$      &$24\pm4$  \\
SFR/pc$^2$            &$\mathrm{M}_{\sun} \, \mathrm{Gyr}^{-1}\pc^{-2}$
                                         &$0.6\pm0.1$   &$3.0\pm0.6$ \\
\hline
\end{tabular}
\end{table}

At a fixed frequency the synchrotron lifetime of the CRE scales with
$\Btot^{~-1.5}$ (Eq.~4), thus for the diffusion length we have $l
\propto (\Btur/\Bord)^{-1} \, \Btot^{~-0.75}$. Therefore, we expect
that in M\,33 $l_\mathrm{x,y}$ is $0.37\pm0.13$ times that in M\,31,
i.e. $l_{x,y}=600\pm230\pc$, but this is about half the observed
value of  $l_\mathrm{x,y}=1270\pm160\pc$. Similarly, the observed
value of $l_\mathrm{z}$ in the thin disk is about twice as large as
expected from scaling with the value in M\,31. As explained in
Sect.~6.1, these larger values may result from the superposition of
the disk and halo emission from M\,33. Thus the superposition further
increases (decreases) the upper limits (lower) limits obtained by
method 1 (see Sect.~\ref{cre}).

We conclude that the differences in $D_\mathrm{Ex,y}$, $l_\mathrm{x,y}$
and $l_\mathrm{z}$ between the thin disks of M\,31 and M\,33 may be
qualitatively understood as the combined effect of the different magnetic
field structures in the two galaxies and the existence of a thick disk /
halo in M\,33.

We note that this is the first time that (a) diffusion lengths and
diffusion coefficients of CRE in the thin disks of galaxies are
derived based on the distributions of the NTH emission alone, and
(b) it is shown that these quantities depend on the properties of
the magnetic field as well as on the occurrence of convective transport
of young CRE from bright SF regions into a thick disk / halo.

\subsection{Influence of star formation}

Above we have seen that M\,31 has no significant vertical magnetic
field and that most of the CRE are confined to the thin disk
(Sect.~5.1). In M\,33, however, CRE are leaving the bright SF regions
along the vertical magnetic field into the thick disk / halo (Sect.~5.2).

Vertical magnetic fields are common in galaxies with thick radio
disks (e.g. \cite{Krause09, Heesen09}) and are often related with
star forming regions. \cite{Dahlem95} first pointed out that thick
radio disks occur in galaxies with a high star formation rate (SFR).
Do M\,31 and M\,33 fit into this picture?

M\,33 indeed has a higher SFR than M\,31 \citep{Verley09,
Tabatabaei10}. We have calculated the mean SFR/pc$^2$ for the areas
under consideration from the extinction-corrected H$\alpha$ maps,
using the same procedure as \cite{Tabatabaei10}. The values given
in Table~6 show that the SFR/pc$^2$ in M\,33 is 5 times higher
than in the bright ring of M\,31. Therefore, the mean TH and NTH
surface brightnesses in M\,33 are higher than in M\,31 (see Table~6;
Fig.~1) because it has more massive stars and SNRs per unit area,
and a stronger total magnetic field.

\cite{Rossa03} analysed the occurrence of gaseous halos observed in
H$\alpha$ using a sample of 74 normal spiral galaxies seen edge-on.
They found that all galaxies with a SFR/pc$^2$ $> (1.4\pm0.2)$
M$_{\sun}$ Gyr$^{-1}$ pc$^{-2}$ show extra-planar diffuse emission
from ionized gas at mean distances of $1-2\kpc$ from the galactic
plane, whereas most galaxies with lower SFR/pc$^2$ only have a thin
disk. According to this division, M\,31 has a thin disk in H$\alpha$
($\mathrm{SFR/pc}^2=(0.6\pm0.1)$ M$_{\sun}$ Gyr$^{-1}$ pc$^{-2}$)
and M\,33 has a thick disk or halo ($\mathrm{SFR/pc}^2=(3.0\pm0.6)$
M$_{\sun}$ Gyr$^{-1}$ pc$^{-2}$). Indeed, in M\,31 the $|z|$-extent
of the diffuse H$\alpha$ emission is less than 500~pc
\citep{Walterbos94} and, based on a power-spectrum analysis,
 \cite{Combes12} suggested that M~33 has a thick H$\alpha$ disk.
Since galaxies with a thick disk in H$\alpha$
emission that have been observed in the radio regime also have a
thick radio disk \citep{Rossa03}, we may expect that M\,33 has a
thick radio disk or halo. Thus the expectations from H$\alpha$
observations are in full agreement with our results of a thin NTH
disk in M\,31 and a thick NTH disk / halo in M\,33 (Table~5).

The SFR/area weakly influences the strength and structure of the
magnetic field in a galaxy. A higher SFR/area generally increases
the total and turbulent magnetic field with (SFR/area)$^{0.2-0.4}$,
but hardly affects the ordered magnetic field (e.g. \citep{Chyzy08,
Krause09, Tabatabaei13a}). This explains why M\,33 has a stronger $\Btur$
than M\,31. The ordered magnetic field, however, is determined by various
dynamical and environmental effects caused by e.g. galactic
rotation, density waves, shear and anisotropic compressions in
spiral arms \citep{Fletcher11}. Although within a galaxy
$\Btur/\Bord$ may increase with (SFR/area)$^{0.3}$ \citep{Chyzy08},
the many factors influencing $\Bord$ may be the reason that between
galaxies the ratio $\Btur/\Bord$ seems independent of SFR
\citep{Krause09}. For example, M\,51 has a 4 times higher SFR/pc$^2$
than M\,33 and strong magnetic fields, but the ratio $\Btur/\Bord$
is similar to that in M\,31 (1.4 compared to $1.2\pm0.5$, see
\cite{Tabatabaei13b}). The strong density waves in M\,51 may
be responsible for this situation.

We have shown that the differences in diffusion length in the thin
disks of M~31 and M~33 are the result of both the different properties
of the magnetic field in the two galaxies and the existence of a thick
disk / halo in M~33, which is absent in M~31. As the field properties
not only depend on the SFR/area, a clear correlation between diffusion
length in the thin disk and SFR/area is not expected. However, galaxies
with SFR/area $> 1.4$ M$_{\sun}$ Gyr$^{-1}$ pc$^{-2}$ are likely to have
a thick radio disk / halo. Then the estimated ``diffusion'' lengths will be
larger than expected from random walk diffusion alone because they are the
result of a combination of diffusion in the thin disk and propagation of
CRE into and inside the halo (Sect.~5.2), and the superposition of disk and
halo onto the sky plane (Sect.~6.1).

\section{Comparison with earlier work}

\subsection{Smearing experiments}

A number of authors estimated the diffusion length of CRE in nearby
galaxies by smearing the distribution of the FIR emission until it
looked similar to that of the emission at 1.4~GHz (e.g.
\cite{Bicay89, Marsh95, Marsh98, Murphy06a, Murphy06b}). However,
these studies have several drawbacks: (1) IR images at $24-70\mum$
were used as the source distribution of the CRE. Although IR
emission is usually well correlated with H$\alpha$ emission within
galaxies (e.g. \cite{Leroy09, Verley09, Tabatabaei10}), the
influence of other dust heating sources than massive stars and of
the dust distribution are not known. (2) Part of the total power
emission observed at 1.4~GHz is thermal emission and its fraction
varies across a galaxy. The inclusion of thermal emission leads to
diffusion lengths that are too small by an amount depending on the
thermal fraction and its distribution in the galaxy. \citet{Murphy08}
did correct the total power emission for the thermal contribution,
but they assumed 2 populations of CRE sources with different mean ages:
one of young SNRs on scales $\lesssim 1$\,kpc and a diffuse one on
scales $>1$\,kpc where CRE are accelerated by interstellar shocks
in the diffuse ISM. We think that there is insufficient evidence for
diffuse CRE sources. (3) The role of magnetic fields was not taken
into account.  (4) The effect of a halo on the derived diffusion
lengths was not considered.

Our study is free of these problems because it is based on properly
separated thermal and nonthermal emission distributions across M\,31
and M\,33 (see Sect.~2). Moreover, the thermal distribution
represents the source distribution of the CRE, and the magnetic
field properties are known. Although the diffusion lengths on the
sky of about 1~kpc in M\,31 and $1-2\kpc$ in M\,33 (Table~4) are in
the range of $0.5-3\kpc$ derived from radio(1.4~GHz) -- FIR studies
(e.g. \cite{Murphy06b}), our values are more reliable estimates of
$l_\mathrm{sky}$.

\cite{Murphy06b, Murphy08, Murphy12} found a steep decrease of their
``diffusion length'' with increasing SFR/area. We note that all galaxies
in their sample have a SFR/area $>(1.4\pm0.2)$ M$_{\sun}$ Gyr$^{-1}$
pc$^{-2}$, indicating that all these galaxies are expected to have a
thick radio disk / halo (see Sect.~6.2). Therefore, their ``diffusion
length'' $l$ includes CRE propagation into the halo and inside the
halo parallel to the disk, making their values larger than expected
for diffusive propagation (Sect.~5.2 and Sect.~6.1). The authors
could not find a mechanism that could explain the
steep slope of their $l$ -- SFR/area relation, possibly
because they neither included the important factor $\Btur/\Bord$ nor the
effect of a thick disk / halo on the observed propagation lengths in their
analysis. The strong dependence of their ``diffusion length'' on SFR/area
may reflect the increasing importance of convection of CRE into a thick
disk / halo by winds from the SF regions and increasing escape of CRE.

\subsection{Implications for the radio -- FIR correlation within
galaxies}

Our study of the radio -- FIR correlations in M\,31 and M\,33
(Fig.~1, Tables~1, 2) confirms that TH emission is about linearly
proportional to emission from warm dust (\cite{Xu92, Hoernes98,
Hippelein03}) and that the correlation between NTH emission at
1.4~GHz and dust emission follows a power law with exponent $b<1$
\citep{Hoernes98}. We have shown that the smaller value of $b$ in
M\,33 compared to that in M\,31 can be explained by the seemingly
larger propagation length of the CRE in M\,33 than in M\,31. Galaxies
with a thick radio disk / halo are expected to have smaller values of $b$
because the combined effect of CRE diffusion in the thin disk and
propagation of CRE into and inside the thick disk / halo leads to a
smoother NTH distribution than diffusion in the thin disk alone.
The superposition of disk and halo emission onto the sky further
smoothens the emission that is observed.

In this context, the work of \cite{Dumas11} on the galaxy M\,51 is
very interesting. They made pixel-to-pixel correlations between dust
emission at $24\mum$ and 1.4~GHz radio continuum emission for
different regions in M\,51: central disk, spiral arms, interarms and
outer parts. Eye estimates of bisector slopes \footnote{Dumas et al.
fitted y(x)-regression lines instead of bisectors.} in their Fig.~11
yield for the exponent $c$ in $I_{1.4\GHz } \propto
I_{24\mum}^{~\mathrm{c}}$ the values $c=0.4$ for the central region,
$c=0.7$ for the spiral arms and $c\simeq1$ for interarm and outer
regions. Since the total 1.4~GHz emission contains thermal emission
with $c=1$, the values of $c<1$ for the central region and the
spiral arms imply that the exponent $b$ in the power law between
nonthermal emission and dust emission, $NTH_{1.4\GHz} \propto
I_{24\mum}^{~b}$, is even smaller than $c$, i.e. $b<0.4$ for the
central region and $b<0.7$ for the arms. \cite{Berkhuijsen97} and
\cite{Fletcher11} showed that M\,51 has a thick NTH disk / halo
extending to a radius of $>3\arcmin$ from the centre. The
high SFR of M\,51 \citep{Leroy09} favours the  propagation of CRE
into and inside the thick disk / halo, which will influence the
observed propagation length. The high magnetic field strengths of
$15-30\muG$ \citep{Fletcher11} yield a mean free path of the CRE or
diffusion length in the thin disk of a few hundred pc \citep{Tabatabaei13b},
which means that few CRE will move
into the interarm and outer regions. In these regions, where $c=1$,
also $b=1$ suggesting that $I_\mathrm{NTH,1.4\,GHz}$ and
$I_{24\mum}$ are directly proportional.

\section{Summary}

We performed classical correlations between the thermal (TH) and
nonthermal (NTH) components of the radio continuum emission at 21~cm
(1.4~GHz) and one higher frequency and far infrared (FIR) dust
emission from M\,31 and M\,33. In both galaxies, the TH emission is
linearly correlated with the emission from warm dust ($24\mum ,\
70\mum$), but the power laws of the NTH -- FIR correlations have
exponents $b<1$ (Fig.~1; Tables~1, 2). The latter increase with
increasing frequency; since the diffusion length of cosmic ray
electrons (CRE) gets smaller towards higher frequencies, this is a
direct indication that the propagation of CRE away from their places
of origin causes $b<1$. The power-law exponents of the
NTH$_{1.4\GHz}$ -- FIR correlations for M\,33 are significantly
lower ($\simeq0.4$) than those for M\,31 ($\simeq0.6$), suggesting
that the propagation length of the CRE in M\,33 is larger than in
M\,31.

We estimated the diffusion length of the CRE in two ways: (1) by
smoothing the NTH emission at the higher frequency until the correlation
with the NTH emission at 1.4~GHz had the power-law exponent $b=1$, and
(2) by smoothing the TH emission until the correlation with the NTH
emission at the same frequency had $b=1$, assuming that the TH emission
represents the source distribution of the CRE, as is the case in M\,33
(Fig.~2). Method 1 yields an estimate of the diffusion length of
the  older CRE within the thin disk of the galaxy and method 2
estimates the propagation length of the bulk of CRE (i.e. of
all ages) resulting from the combination of diffusion in the thin
disk and convective propagation (galactic wind) of CRE out of
the thin disk into and within a thick disk / halo.

Our estimates for the propagation lengths of CRE are the first ones
obtained from the distributions of the NTH components of the radio emission.

Our main results are summarized as follows:

1. In M\,31 most of the CRE are confined to a thin NTH disk with
an exponential scale height of $0.3-0.4\kpc$ at 1.4~GHz, whereas in
M\,33 many CRE move out of the thin disk into a thick disk / halo
with scale height $h_\mathrm{thick}\simeq2\kpc$ at 1.4~GHz. A thick
NTH disk or halo in M\,33 (but not in M\,31) is consistent with the
higher SFR in M\,33 and the existence of a vertical magnetic field
component.

2. The propagation of CRE from the SF regions into and inside the
thick disk / halo, and the superposition of disk and halo emission
on the sky lead to larger apparent propagation lengths in M\,33 than
the diffusion lengths in M\,31 along and perpendicular to the disk plane
(method 2).

3. We showed that in case of random walk diffusion in the thin
disk $l_\mathrm{x,y}\propto (\Btur/\Bord)^{-1} \, \Btot^{~-0.75}$ is
expected (Sect.~\ref{structure}), leading to a 3 times smaller value
in M\,33 than in M\,31. However, the (deprojected) diffusion length in
M\,33 is only 30\% smaller than that in M\,31, i.e.
$l_\mathrm{x,y}=1.3\pm0.2\kpc$ and $l_\mathrm{x,y}=1.6\pm0.3\kpc$ at
1.4~GHz, respectively (Table~5). The high value of $l_\mathrm{x,y}$
in M\,33 is due to the thick disk / halo emission seen superimposed
onto the disk emission on the sky.

4. For galaxies with a thick disk / halo the apparent propagation length
derived from smearing experiments may be considerably larger than the actual
diffusion length in the thin disk.

We conclude that the spatial structure, strength and regularity
of the magnetic field in a galaxy as well as the occurrence of a galactic
wind causing a thick disk / halo determine the propagation of the CRE.
Hence, the power-law exponent of the NTH -- FIR correlation within a galaxy
is the result of the properties of the magnetic field and the existence of a
thick disk / halo.

\section*{Acknowledgments}

We thank Gabi Breuer \footnote{who sadly passed away on 2013 May 29th} for
getting the tables in the correct format. We thank Marita Krause  for useful
comments on the manuscript. Critical comments of the referee, Andrew Strong,
lead to a better description of the basic assumptions in the paper.

\bibliographystyle{mn2e}

\bibliography{m31m33cre}

\end{document}